\documentclass[a4paper,11pt]{article}
\usepackage{pos}

\usepackage{bm}
\usepackage{xcolor}
\usepackage{cleveref}

\usepackage[T1]{fontenc} 

\crefname{section}{section}{sections}
\Crefname{section}{Section}{Sections}

\crefname{figure}{figure}{figures}
\Crefname{figure}{Figure}{Figures}
\crefrangeformat{figure}{figures~#3#1#4--#5#2#6}
\Crefrangeformat{figure}{Figures~#3#1#4--#5#2#6}

\crefrangeformat{equation}{eqs.~(#3#1#4)--(#5#2#6)}
\Crefrangeformat{equation}{Equations~(#3#1#4)--(#5#2#6)}

\creflabelformat{subequations}{(#2#1#3)}
\crefname{subequations}{eqs.}{eqs.}
\crefrangeformat{subequations}{eqs.~(#3#1#4)--(#5#2#6)}
\Crefname{subequations}{Equations}{Equations}
\Crefrangeformat{subequations}{Equations~(#3#1#4)--(#5#2#6)}

\newcommand{\as}{\ensuremath{\alpha_\text{s}}}

\newcommand{\ag}{\ensuremath{\alpha_\text{g}}}

\newcommand{\dd}{\ensuremath{\mathrm{d}}}
\newcommand{\e}{\ensuremath{\mathrm{e}}}
\newcommand{\iunit}{\ensuremath{\mathrm{i}}}
\newcommand{\case}[2]{\ensuremath{{\textstyle\frac{#1}{#2}}}}
\newcommand{\half}{\ensuremath{\case{1}{2}}}

\newcommand{\MSbar}{\ensuremath{\overline{\rm MS}}}
\newcommand{\MSbbar}{\ensuremath{\overline{\bf MS}}}

\begin{document}

\title{More on minimal renormalon subtraction}
\ShortTitle{More on minimal renormalon subtraction}

\author*{Andreas S. Kronfeld}

\affiliation{Theory Division, Fermi National Accelerator Laboratory,$^\dagger$ \\
    P.O. Box 500, Batavia IL 60565, USA}
\affiliation{Institute for Advanced Study, Technische Universität München, \\
    Lichtenbergstra\ss e~2a, 85748 Garching, Germany}
\notes{\note{Fermilab is managed by Fermi Research Alliance, LLC, under Contract No.\ DE-AC02-07CH11359 with the U.S.\ Department of
Energy.}}

\abstract{The minimal renormalon subtraction (MRS) \cite{FermilabLattice:2018est,Brambilla:2017hcq,Komijani:2017vep,%
Kronfeld:2023jab} technique is summarized.
A new result is a study of the scale dependence of the pole-mass--\MSbar-mass ratio in MRS perturbation theory.
As expected, the scale dependence is much milder than in standard perturbation theory, but it is a bit larger than other truncation 
effects such as omitting the N\textsuperscript{3}LO term or varying the normalization of the renormalon subtraction.}

\FullConference{The 40th International Symposium on Lattice Field Theory (Lattice 2023)\\
July 31st - August 4th, 2023\\
Fermi National Accelerator Laboratory\\}


\maketitle

\section{Introduction}

Many quantities calculated with lattice gauge theory have to be ``matched'' to continuum QCD.
For quantities that run, such as \as\ and quark masses, it is conventional to quote results in the \MSbar\ scheme, making
perturbation theory unavoidable.
The uncertainty introduced by truncating the perturbative series is notoriously difficult to estimate.

It is therefore noteworthy that the error budget for the \MSbar\ quark masses of ref.~\cite{FermilabLattice:2018est} quotes a
negligible uncertainty from perturbative truncation.
Figure~4 of ref.~\cite{FermilabLattice:2018est} demonstrates a small shift in the final results when the order of perturbation
theory changes from NLO to NNLO and an imperceptible shift from NNLO to N\textsuperscript{3}LO (counting the one-loop term as LO).
The total uncertainty is dominated by statistics, fitting systematics, and the external parametric error of~\as.
It ranges from subpercent for bottom, charm, and strange to 1--2\% for down and up.

To tame the perturbative series, which in this case is the relation between the pole mass and the \MSbar\ mass, the key ingredient
was a rearrangement and reinterpretation of the series dubbed ``minimal renormalon subtraction'' (MRS)~\cite{Brambilla:2017hcq},
which in turn relied on an expression for the normalization of factorial growth of the series coefficients~\cite{Komijani:2017vep}.
The MRS method has been generalized and clarified in work underway during Lattice 2023 and meanwhile
published~\cite{Kronfeld:2023jab}.
This report summarizes the main points and shows plots examining the scale dependence of the MRS quark mass.
At the conference, I showed plots of the static energy (the others were not yet ready).
The two sets of plots look very similar.
References~\cite{FermilabLattice:2018est,Brambilla:2017hcq,Komijani:2017vep} did not study how MRS perturbation theory depends on
$s=\mu/\bar{m}$, where $\mu$ is the scale in $\alpha_{\MSbar}(\mu)$, and $\bar{m}=m_{\MSbar}(\bar{m})$.
Showing the quark-mass plots here fills a gap while avoiding rote repetition of ref.~\cite{Kronfeld:2023jab}.
In particular, \cref{fig:err}~(right) suggests that the truncation uncertainty in the quark-mass results of
ref.~\cite{FermilabLattice:2018est} may have been underestimated, but arguably not so much that it would alter the quadrature sum of
all uncertainties.

\section{Statement of the problem}
\label{sec:statement}

References~\cite{FermilabLattice:2018est,Brambilla:2017hcq,Komijani:2017vep} start with a relation for the mass of a heavy-light 
hadron,
\begin{equation}
    M = m_\text{pole} + \bar\Lambda + \frac{\mu_\pi^2}{2m_\text{pole}^2} + \cdots ,
    \label{eq:hqet}
\end{equation}
where $M$ is the hadron mass, $m_\text{pole}$ is the quark ``pole'' mass, $\bar\Lambda$ is the energy of gluons and light quarks,
and the last term is the heavy quark's kinetic energy.
A spin-dependent term of order $1/m_\text{pole}^2$ can also arise; for this discussion, it can be lumped into the last term.
In perturbation theory, the pole mass is obtained from the self-energy by finding the pole of the propagator and substituting
iteratively, because the self-energy itself depends on $m_\text{pole}$~\cite{Kronfeld:1998di}.
The relation can be written
\begin{equation}
    m_\text{pole} = \bar{m}\left(1 + \sum_{l=0}r_l(\mu/\bar{m})\as^{l+1}(\mu)\right) ,
    \label{eq:series}
\end{equation}
and the coefficients grow rapidly: $r_l(1)=\{0.424, 1.035, 3.693, 17.21\}$ for $l=\{0, 1, 2, 3\}$.
(The numbers correspond to the \MSbar\ scheme with $n_f=3$ massless quarks and no massive quark sea.) %

If $m_\text{pole}$ solves the perturbative pole condition, so does $m_\text{pole}-\delta m$, where $\delta m$ is anything of order
$\Lambda\sim\mu\e^{-1/2\beta_0\as(\mu)}$ (the QCD scale).
In the hadron mass, $\delta m$ can be absorbed into $\bar\Lambda$, which isn't defined without specifying exactly what
$m_\text{pole}$ means in \cref{eq:hqet}.
This ambiguity, often called the ``renormalon'' ambiguity~\cite{Bigi:1994em,Beneke:1994sw}, is related to the growth of the
coefficients, which is factorial.
Indeed, the MRS procedure introduces some rearrangements and eventually snips off a $\delta m$ (given below, \cref{eq:deltam}),
absorbing it into $\bar\Lambda$ to define $\bar{\Lambda}_\text{MRS}$.
Then $M=m_\text{MRS}+\bar{\Lambda}_\text{MRS}+\cdots$.

There are many problems of the form \cref{eq:hqet}.
Generically,
\begin{equation}
    Q^p \mathcal{R}(Q) = Q^p R(Q) + C_p\Lambda^p + \cdots , \qquad
    R(Q) = \sum_{l=0}r_l(\mu/Q)\as^{l+1}(\mu) ,
    \label{eq:QR}
\end{equation}
where $\mathcal{R}$ is some dimensionless observable with one short-distance scale~$Q$.
In the following, formulas are written for arbitrary $p$~\cite{Kronfeld:2023jab}, although we are mostly interested in $p=1$.

\section{Factorial growth}
\label{sec:bang}

It has been known for a long time (see, e.g., ref.~\cite{Beneke:1998ui}) that the coefficients in the series of
\cref{eq:series,eq:QR} grow at \emph{large} orders as
\begin{subequations} \label[subequations]{eq:oldRlR0}
\begin{equation}
    r_l \sim \left(\frac{2\beta_0}{p}\right)^l \frac{\Gamma(l+1+pb)}{\Gamma(1+pb)} R_0 \equiv R_l
    \label{eq:oldRl}
\end{equation}
where $b=\beta_1/2\beta_0^2\stackrel{n_f=3}{=}32/81\approx0.4$.
Komijani~\cite{Komijani:2017vep} found an expression for the $l$-independent normalization $R_0$
\begin{equation}
    R_0 = \sum_{k=0}^{\textcolor{purple}{\infty}} (k+1) \frac{\Gamma(1+pb)}{\Gamma(k+2+pb)} \left(\frac{p}{2\beta_0}\right)^k 
        f_k ,
    \label{eq:oldR0}
\end{equation}
\end{subequations}
where the coefficients $f_k$ are straightforwardly related to the $r_l$ (cf.\ \cref{eq:fk}, below).
Komijani observed that since $\bar\Lambda$ in \cref{eq:hqet} (more generally, $C_p\Lambda^p$ in \cref{eq:QR}) does not depend on
$\bar{m}$ (or $Q$ in \cref{eq:QR}~\cite{Kronfeld:2023jab}), a derivative with respect to $\bar{m}$ (or $Q$) yields a series without
the leading factorial growth; the coefficients of that series are the~$f_k$.
In \cref{eq:oldR0}, the upper limit $\textcolor{purple}{\infty}$ is an asymptotic consideration.
If $L$ of the $r_l$ are known, then only $L$ of the $f_k$ are known.
Thus, any practical application must truncate the sum in \cref{eq:oldR0} to have upper limit $\textcolor{purple}{L-1}$.

An unexpected---perhaps remarkable---finding~\cite{Kronfeld:2023jab} is a similar set of formulas that hold at every order, not just
large orders.
My version reads
\begin{equation}
    r_l = f_l + \left(\frac{2\beta_0}{p}\right)^l \frac{\Gamma(l+1 + pb)}{\Gamma(1+pb)}
        \sum_{k=0}^{\textcolor{purple}{l-1}} (k+1) \frac{\Gamma(1+pb)}{\Gamma(k+2 + pb)} \left(\frac{p}{2\beta_0}\right)^k f_k .
    \label{eq:rrenorm}
\end{equation}
Note that the relation is $=$ (equals) not $\sim$ (asymptotically goes like).
Gaze at \cref{eq:rrenorm} and see that the growing part (in front of the sum) is exactly like \cref{eq:oldRl} and that the sum 
itself is almost like the normalization, \cref{eq:oldR0}, albeit with upper limit the depending on $l$.

\Cref{eq:rrenorm} is nothing but the inverse of the formula for the $f_k$ in a carefully chosen scheme.
In any scheme for \as,
\begin{equation}
    f_k = r_k-\frac{2}{p}\sum_{l=0}^{k-1} (l+1) \beta_{k-1-l} r_l .
    \label{eq:fk}
\end{equation}
The algebra can be simplified by changing from the \MSbar\ scheme, which is used in the literature to report the
$r_l$~\cite{Gray:1990yh,Chetyrkin:1999ys,Chetyrkin:1999qi,Marquard:2015qpa,Marquard:2016dcn}, to a scheme with
$\beta_j=\beta_0(\beta_1/\beta_0)^j$.
With these $\beta_j$, the $\beta$-function can be summed as a geometric series, so this scheme is known as the ``geometric scheme''.
Without loss, it is possible to define the coupling in this scheme, \ag, such that $\Lambda_\text{g}=\Lambda_{\MSbar}$.
\Cref{eq:fk} implies a matrix that in the geometric scheme has a simple recursive structure.
The matrix has infinite extent but is lower triangular and can be inverted row-by-row.
The recursive nature of the matrix elements yields the $\Gamma$ functions in \cref{eq:rrenorm}.

Has anything been gained?
After all, results are published for the $r_l$ and the $f_k$ are obtained (after deciding what $p$ should be) from \cref{eq:fk}.
For the first $L$ terms, \cref{eq:rrenorm} simply returns the original information.
For $l\ge L$, however, \cref{eq:rrenorm} suggests an approximation as follows.
For the first $L$ coefficients $r_l$, use the published information.
For the rest, namely $r_l$ with $l\ge L$, truncate \cref{eq:rrenorm}, which amounts to dropping the first term and using
\cref{eq:oldRlR0} with upper limit on the sum in $R_0$ at $L-1$.
The truncation is systematic, once $l$ is large enough, because both $f_l$ and any terms omitted from the sum in \cref{eq:rrenorm}
are factorially smaller than the terms that are retained.

\section{Reinterpreting the series}
\label{sec:Borel}

\subsection{Approach of ref.~\cite{Kronfeld:2023jab}}

In the approximation just described, the series is approximated with
\begin{equation}
    R(Q) = \sum_{l=0}^{L-1} r_l \alpha^{l+1} + \sum_{l=L}^\infty R_l \alpha^{l+1} ,
    \label{eq:approx}
\end{equation}
using the $L$ terms in the literature~\cite{Gray:1990yh,Chetyrkin:1999ys,Chetyrkin:1999qi,Marquard:2015qpa,Marquard:2016dcn} and
approximating the rest from the obvious truncation of \cref{eq:rrenorm}.
If the lower limit of the second sum is lowered to $0$, then the sum can be defined via Borel summation.
To add those $L$ terms in, they also have to be subtracted out, i.e.,
\begin{equation}
    R(Q) = \sum_{l=0}^{L-1} [r_l-R_l] \alpha^{l+1} + \sum_{l=0}^\infty R_l \alpha^{l+1}.
    \label{eq:approx-rearrange}
\end{equation}
Then (see ref.~\cite{Kronfeld:2023jab} for details)
\begin{equation}
    \sum_{l=0}^\infty R_l \ag^{l+1} = R_0 \frac{p}{2\beta_0} \mathcal{J}(pb,1/2\beta_0\ag) + \delta R,
    \label{eq:Borel}
\end{equation}
where $\mathcal{J}$ has a convergent expansion in its second argument, i.e., in $1/\ag$.
The last term in \cref{eq:Borel} is explicitly of order $\Lambda_{\MSbar}^p$, so it can be absorbed into the power term in
\cref{eq:QR} (or $\bar\Lambda$ in \cref{eq:hqet}).
For example, for the pole mass (where $p=1$)
\begin{equation}
    \delta m = \bar{m}\,\delta R = - R_0 \e^{\pm\iunit b\pi} \frac{\Gamma(-b)}{2^{1+b}\beta_0}
        \left[\bar{m}\frac{\e^{-1/[2\beta_0\ag(\bar{m})]}}{[\beta_0\ag(\bar{m})]^b}\right] ,
    \label{eq:deltam}
\end{equation}
where the sign in $\e^{\pm\iunit b\pi}$ exposes an ambiguity (from a choice of contour in a Borel integration), and the quantity in
brackets [\;] is immediately recognizable as $\Lambda_\text{g}=\Lambda_{\MSbar}$.

In summary, the ``minimal renormalon subtraction'' (MRS) interpretation of a perturbative series is
\begin{equation}
    R(Q) \to R_\text{MRS}(Q) \equiv \sum_{l=0}^{L-1} [r_l-R_l] \alpha^{l+1} + R_0 \frac{p}{2\beta_0} \mathcal{J}(pb,1/2\beta_0\ag).
    \label{eq:RMRS}
\end{equation}
For the pole mass, $r_l-R_l=\{-0.096, -0.004, 0.128, -0.127\}$ (at $\mu=\bar{m}$), much smaller than the $r_l$ given in
\cref{sec:statement}.
For $l=3$, for example, $0.127$ is much smaller than $17.2$.

\subsection{Contrast with ref.~\cite{Brambilla:2017hcq}}
\label{sec:synopsis}

The reasoning given in ref.~\cite{Brambilla:2017hcq} is a bit different.
If the formulas hold only for asymptotically large orders, it is a prescription to add and subtract them and a practical concession 
to truncate:
\begin{equation}
    \sum_{l=0}^\infty r_l \ag^{l+1} \to \sum_{l=0}^\infty [r_l-R_l] \ag^{l+1} +
        \sum_{l=0}^\infty R_l \ag^{l+1} \to
        \sum_{l=0}^{L-1} [r_l-R_l] \ag^{l+1} + R_0 \frac{p}{2\beta_0} \mathcal{J}.
        \label{eq:PTjavad}
\end{equation}
The reasoning of ref.~\cite{Kronfeld:2023jab} is that medium and large orders are introduced in a well-founded approximation, as
sketched in \cref{eq:approx,eq:approx-rearrange,eq:Borel,eq:RMRS}:
\begin{equation}
    \sum_{l=0}^\infty r_l \ag^{l+1} \to \sum_{l=0}^{L-1} r_l \ag^{l+1} +
        \sum_{l=L}^\infty R_l \ag^{l+1} \to
        \sum_{l=0}^{L-1} [r_l-R_l] \ag^{l+1} + R_0 \frac{p}{2\beta_0} \mathcal{J}.
        \label{eq:PTandreas}
\end{equation}
Standard, fixed-order perturbation theory is merely
\begin{equation}
    \sum_{l=0}^\infty r_l \ag^{l+1} \to \sum_{l=0}^{L-1} r_l \ag^{l+1} ,
        \label{eq:PTstandard}
\end{equation}
the poor convergence of which is the \emph{raison d'être} for more sophisticated approaches.

\section{Scale-dependence of the pole-mass--\texorpdfstring{\MSbbar}{MS}-mass relation}
\label{sec:mass-study}

This section shows results for MRS applied to the quark mass.
As noted in the introduction, at the conference I showed preliminary versions of plots for the static energy, which have been
finalized and published~\cite{Kronfeld:2023jab}.
It is more interesting to show the quark-mass plots here, in particular to fill a gap in the information in 
refs.~\cite{FermilabLattice:2018est,Brambilla:2017hcq,Komijani:2017vep}.
That work evaluated $\as(\mu)$ and the corresponding coefficients at $\mu=\bar{m}$ without examining other choices.
Here, $\mu=s\bar{m}$ is considered, with the plots showing the variation for $s\in\{\half,1,2\}$.

To be specific, this section shows results for a massive valence quark with three massless sea quarks.
In refs.~\cite{FermilabLattice:2018est,Brambilla:2017hcq}, the charmed sea was included albeit decoupled, following
ref.~\cite{Ayala:2014yxa}.
For simplicity, the charmed sea is omitted here.

\begin{figure}
    \includegraphics[width=0.48\textwidth]{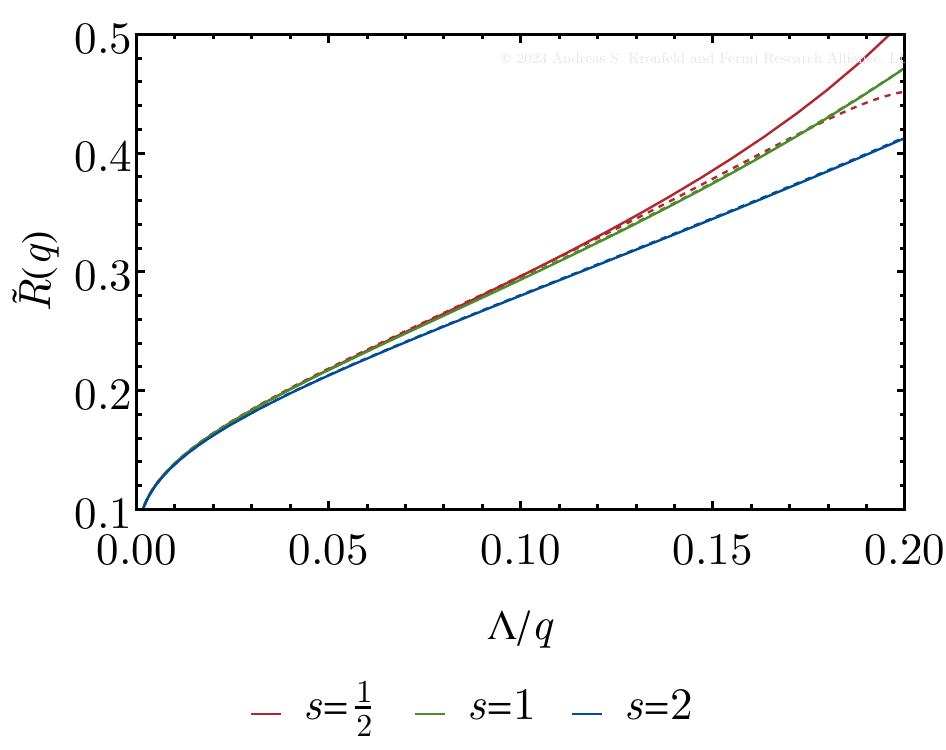}\hfill
    \includegraphics[width=0.48\textwidth]{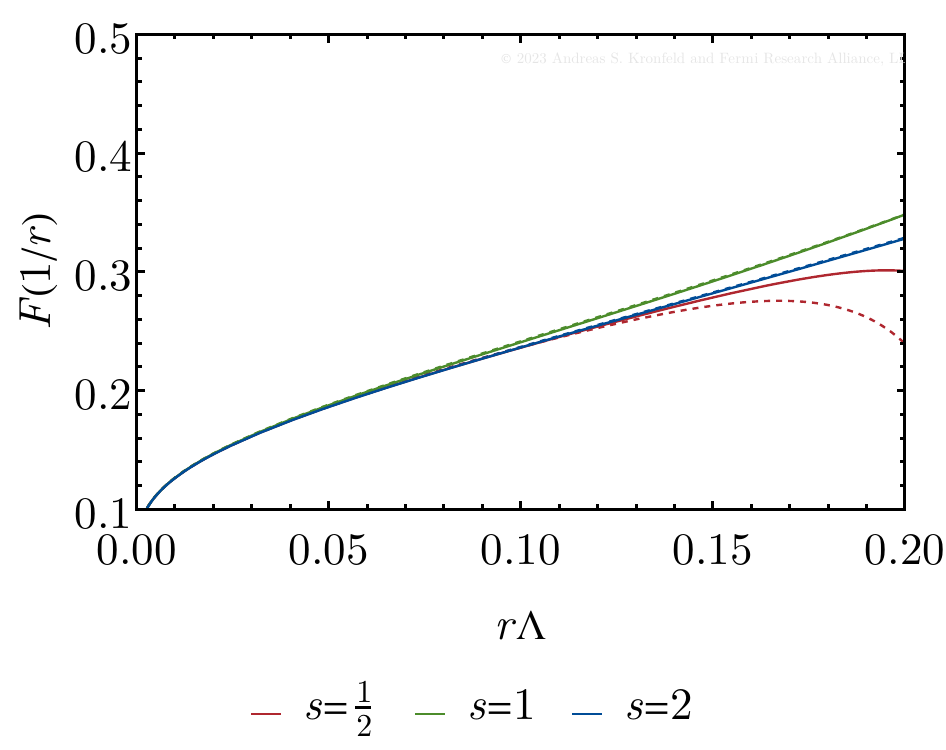}
    \caption[fig:good]{Left: momentum-space static energy $\tilde{R}(q)=-q^2\tilde{V}(q)/C_F$ vs.\ $\Lambda/q$ for $\mu=sq$; right:
        coordinate-space static force $F(r)=(r^2/C_F)\dd V/\dd r$ vs.\ $r\Lambda$ for $\mu=s/r$; in both cases, $s\in\{\half,1,2\}$.
        From ref.~\cite{Kronfeld:2023jab}.}
    \label{fig:good}
\end{figure}

For orientation, \cref{fig:good} shows examples of a good series and an excellent series.
The details are in ref.~\cite{Kronfeld:2023jab}.
The important feature here is how little variation with $s$ arises with four nontrivial terms in the series.
\begin{figure}
    \includegraphics[width=0.48\textwidth]{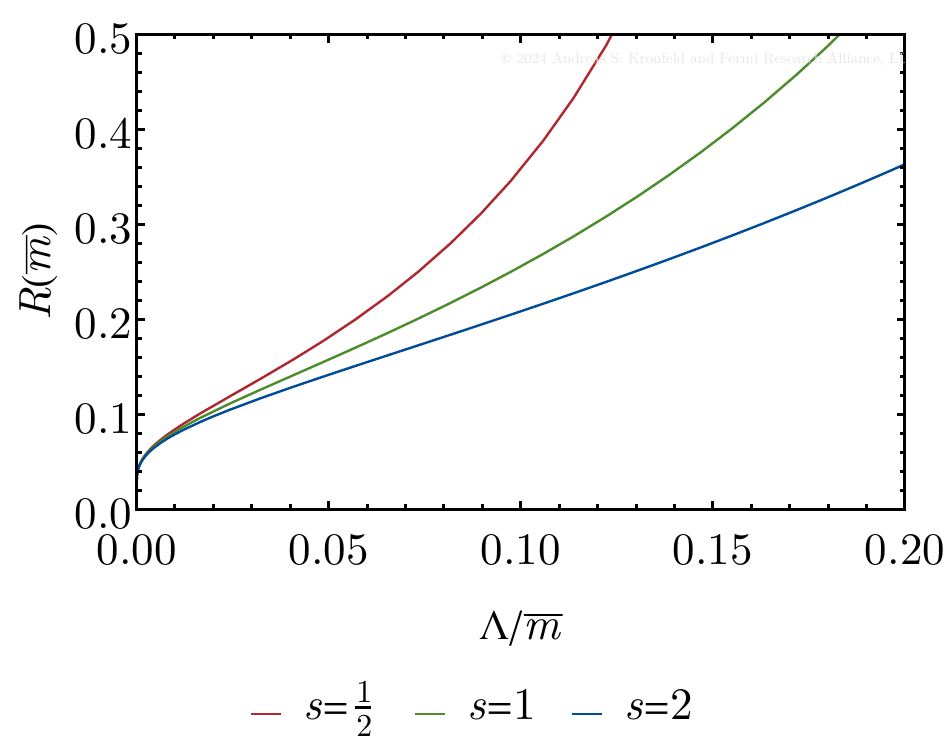}\hfill
    \includegraphics[width=0.48\textwidth]{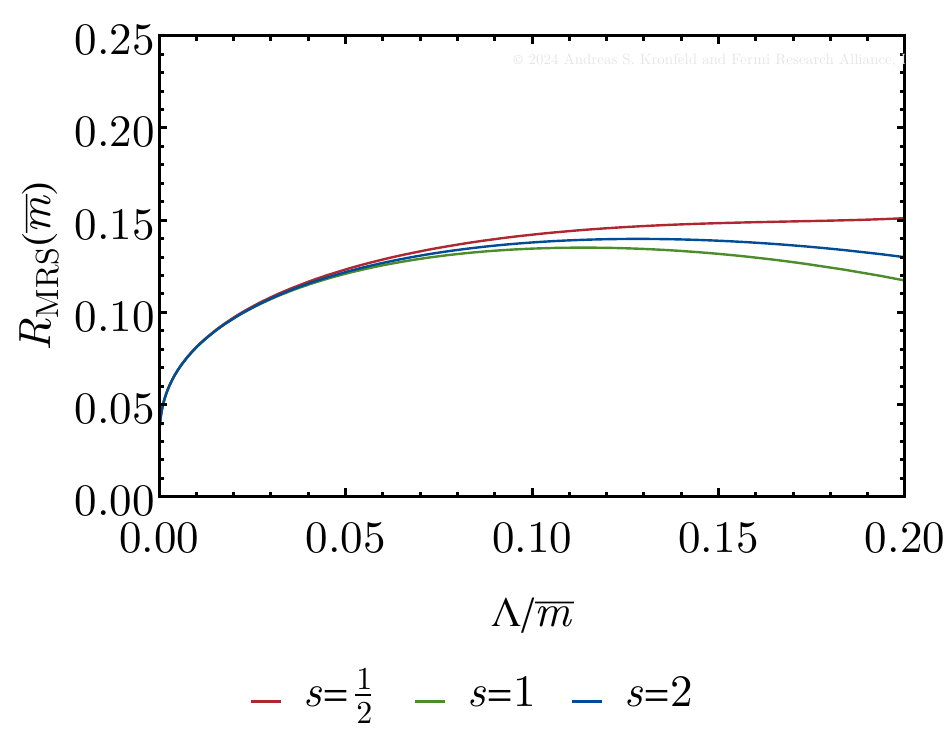}
    \caption[fig:good]{Left: fixed-order perturbation theory for $R(\bar{m})\equiv m_\text{pole}/\bar{m}-1$ vs.\ $\Lambda/\bar{m}$
        for $\mu=sq$; right: MRS perturbation theory for $R(\bar{m})$ for $\mu=s/r$; in both cases, $s\in\{\half,1,2\}$.}
    \label{fig:pole}
\end{figure}
Fixed-order ($L=4$) and MRS definitions of the series are shown in \cref{fig:pole}.
The fixed-order estimate of $R(\bar{m})\equiv m_\text{pole}/\bar{m}-1$ is a disaster, but the MRS prescription exhibits a variation
with the choice of $s$ that is as mild as the (renormalon-free) static force.
As with the static energy, the subtracted, truncated series and the Borel sum on the right-hand side of \cref{eq:RMRS} both vary 
significantly with $s$, but the total does not.

\begin{figure}
    \includegraphics[width=0.48\textwidth]{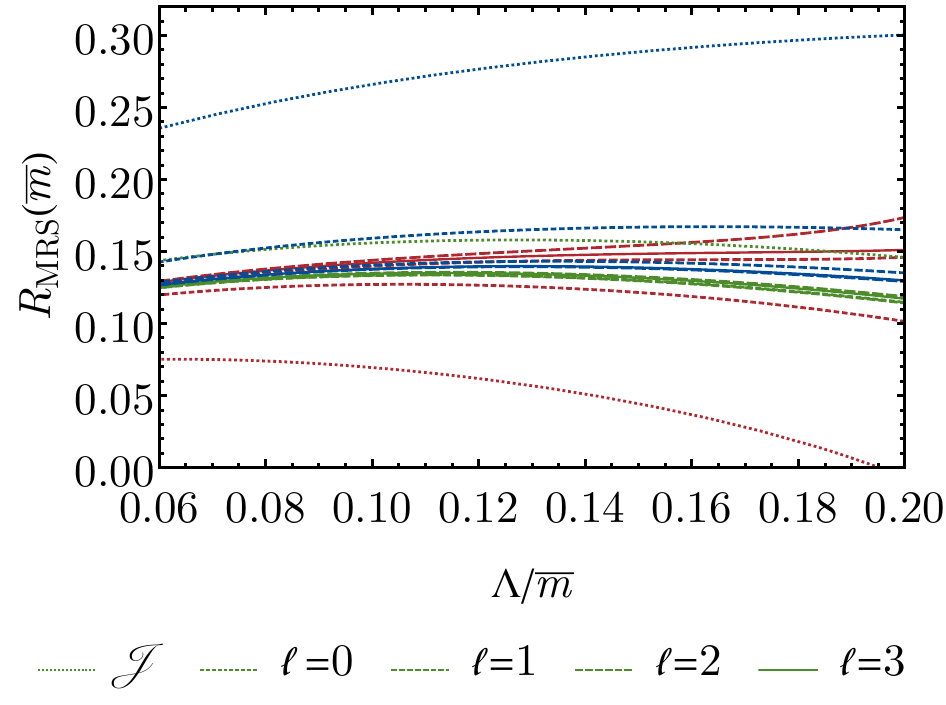}\hfill
    \includegraphics[width=0.48\textwidth]{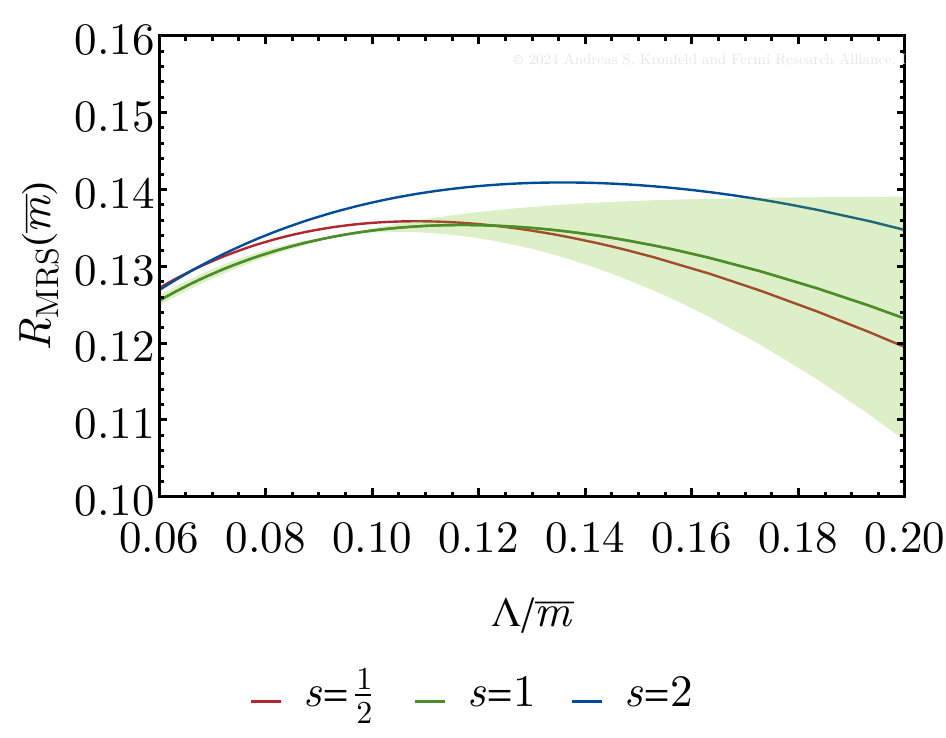}
    \caption[fig:good]{Left: accumulating MRS perturbation theory order-by-order for $R(\bar{m})\equiv m_\text{pole}/\bar{m}-1$
        vs.\ $\Lambda/\bar{m}$ for $\mu=sq$.
        Right: the uncertainty on $R_0$ propagated to $R_\text{MRS}$; here $\mu^2=(2~\text{GeV})^2+(s\bar m)^2$.}
    \label{fig:err}
\end{figure}

\Cref{fig:err}~(left) shows the convergence of MRS perturbation theory for the pole-mass--\MSbar-mass ratio.
The color code is them same as in \cref{fig:good,fig:pole}, and the order is indicated by the length of the dashes, with dotted
curves show the $\mathcal{J}$ term only and solid showing the full N\textsuperscript{3}LO result.
The overlapping curves show that the changes made for each term are small for NLO, very small for NNLO, and extremely small for
N\textsuperscript{3}LO.

In refs.~\cite{FermilabLattice:2018est,Brambilla:2017hcq,Komijani:2017vep}, the uncertainty in the normalization of the factorial 
terms, $R_0$, is estimated to be the size of the last term, which in pole mass as for the static energy is the fourth term.
\Cref{fig:err}~(right) shows how this error estimate propagates to $R_\text{MRS}$.
Superficially, it is 2--4 times less than the uncertainty from scale variation.
That means the truncation uncertainty in the quark-mass results~\cite{FermilabLattice:2018est} may be a bit larger than thought.
After refitting the data (not yet done), it seems unlikely that it would become larger than the other, dominant uncertainties.

%

\bibliographystyle{JHEP}
\bibliography{j}

\end{document}